\documentclass{article}
\usepackage{booktabs}
\usepackage{cite}
\usepackage{diagbox}
\usepackage{subfigure}

\usepackage[pdftex]{graphicx}

\usepackage[cmex10]{amsmath}
\usepackage{amssymb}

\title{Joint Transceiver Optimization for Wireless Communication PHY with Convolutional Neural Network}

\author{Banghua~Zhu,
       Jintao~Wang, 
       Longzhuang~He,
       and~Jian~Song
\thanks{Banghua Zhu, Jintao Wang, Longzhuang He and Jian Song are with the Department
of Electronic Engineering, Tsinghua University, Beijing, 100084 China. E-mail: 13aeon.v01d@gmail.com.}% <-this % stops a space
\thanks{This work was supported in part by the National Key R\&D Program of China (Grant No.2017YFE011230) and the National Natural Science Foundation of China (Grant No.61471221). (Corresponding author: Jintao Wang.) }}

\begin{document}

\maketitle

\begin{abstract}
 Deep Learning has a wide application in the area of natural language processing and image processing due to its strong ability of generalization. In this paper, we propose a novel neural network structure for jointly optimizing the transmitter and receiver in communication physical layer under fading channels. We build up a convolutional autoencoder to simultaneously conduct the role of modulation, equalization and demodulation. The proposed system is able to design different mapping scheme from input bit sequences of arbitrary length to constellation symbols according to different channel environments.  The simulation results show that the performance of neural network based system is superior to traditional modulation and equalization methods in terms of time complexity and bit error rate (BER) under fading channels. The proposed system can also be combined with other coding techniques to further improve the performance. Furthermore, the proposed system network is more robust to channel variation than traditional communication methods.

\end{abstract}

% Note that keywords are not normally used for peerreview papers.

% For peer review papers, you can put extra information on the cover
% page as needed:
% \ifCLASSOPTIONpeerreview
% \begin{center} \bfseries EDICS Category: 3-BBND \end{center}
% \fi
%
% For peerreview papers, this IEEEtran command inserts a page break and
% creates the second title. It will be ignored for other modes.

\section{Introduction}
% The very first letter is a 2 line initial drop letter followed
% by the rest of the first word in caps.
% 
% form to use if the first word consists of a single letter:
% \IEEEPARstart{A}{demo} file is ....
% 
% form to use if you need the single drop letter followed by
% normal text (unknown if ever used by IEEE):
% \IEEEPARstart{A}{}demo file is ....
% 
% Some journals put the first two words in caps:
% \IEEEPARstart{T}{his demo} file is ....
% 
% Here we have the typical use of a "T" for an initial drop letter
% and "HIS" in caps to complete the first word.
% \IEEEPARstart{T}{his} demo file is intended to serve as a ``starter file''
% for IEEE journal papers produced under \LaTeX\ using
% IEEEtran.cls version 1.8a and later.
% % You must have at least 2 lines in the paragraph with the drop letter
% % (should never be an issue)
% I wish you the best of success.

One of the most critical problem in wireless communication techniques is how information gets transmitted from one end (transmitter) to another end (receiver) reliably. During the transmission, the signal suffers from distortion and noise due to complicated channel state and hardware imperfections. In the study of physical layer of the OSI model (PHY), the whole system is optimized in a divide-and-conquer perspective \cite{tse2005fundamentals}. The PHY transmitter usually includes source coding, channel coding, and modulation module, while the PHY receiver includes synchronization, channel estimation, equalization, demodulation, channel decoding, source decoding, and so on.
Each part is optimized separately and requires a great amount of expert knowledge. Massive research has been focused on the optimization of each module for different channel environments and application demands. According to data processing theorem \cite{ziv1973functionals} in information theory, the optimization of sub-modules for communications cannot guarantee global optimality for the whole communication system. In fact, such an implementation is known to be sub-optimal \cite{zehavi19928}.
% We cascade all the modules to form a whole communication system.
 
In the past decades, deep learning has seen wide and successful application in computer vision and natural language processing due to its strong ability of generalization. In order to further improve the transmission and network performance, the fifth generation communication system (5G) will apply many new techniques, such as massive MIMO \cite{driessen1999capacity}, mmWave \cite{rappaport2013millimeter}, and ultra-dense wireless network \cite{ge20165g}. As a result, it also raises a great number of challenges and opportunities. In communication area, deep learning is applied to communication network \cite{wang2017deep} and cognitive radio \cite{bkassiny2013survey}. However, since physical layer is quite complex and requires real time and high accuracy  transmission, the development of related technology has been insufficient until recent years \cite{ibnkahla2000applications}. With the development of neural network compression technique and specialized hardware such as GPU and FPGA, the cost and time complexity of deep learning related techniques are significantly decreased. It is possible to run neural network on mobile devices and antennas. There has been massive research on deep learning based techniques on individual modules such as channel decoding \cite{nachmani2016learning,cammerer2017scaling,gruber2017deep,liang2018iterative,kim2018communication}, signal detection \cite{samuel2017deep,jeon2017blind,hong2017mimo,farsad2017detection}, and etc.

In deep learning based communication systems, it is possible to optimize the transmitter and receiver jointly with a structure of autoencoder \cite{salakhutdinov2000nonlinear} instead of artificially introduced block schemes \cite{o2016learning, o2017introduction,o2017deep}. In previous efforts, the autoencoder is able to automatically design the mapping scheme from an one-hot encoded signal to constellation symbols and the way of demapping under AWGN channel. The network is proven to outperform the traditional coding techniques such as uncoded BPSK and Hamming coded BPSK scheme in the aspect of block error rate \cite{o2017introduction}. However, the application of this work is still limited since the input can only be short one-hot vector, which carries much less information than a random bit sequence of the same length. Also, uncoded BPSK and Hamming coded BPSK scheme is not specially designed for one-hot vector. The training of neural network has a natural advantage over the benchmark. Furthermore, the size of the dense layer based autoencoder would grow with the increased length of the input sequence, which would lead to quadratic growth in time complexity. Such issues as above have not been well addressed by the current research.

In another work on autoencoder communication system \cite{dorner2018deep}, the authors addressed the
issue of receiver synchronization by introducing a frame synchronization scheme based on another neural network. The authors further proposed a transfer learning based two-phase procedure to overcome the problem of missing channel gradient during training. The transmitter and receiver are trained under a stochastic channel model, and the receiver is then finetuned with respect to real channels.  These work enable practical over-the-air transmission through pure neural networks. 

In this paper, we focus on further exploring the application of autoencoder in physical layer. Our main contribution is listed as follows.
\begin{itemize}
\item We propose a novel structure based on convolutional autoencoder which is able to jointly optimize the transmitter and receiver design from overall system performance point of view which has the following virtues: 1. the property of convolutional autoencoder enables the trained network to process input bit sequence of any length, 2. the soft output of this structure can be the input of any soft decoder, which can be easily combined with any other soft-input-soft-output channel decoding schemes, and 3. the proposed structure can be flexibly applied to either time or frequency domains.
\item We have confirmed that the proposed structure is able to handle the challenge of mapping scheme of different levels under various channels, including AWGN channel, fading channels and non-Gaussian noise channels.
\item From various BER performance evaluation and time complexity analysis of the proposed system, the robustness of the system is demonstrated. The proposed system structure is less sensitive to channel variation compared with traditional design, showing the great potential for the design principle and methodology of future telecommunication systems.
\end{itemize}

{\it Notation.}  Bold capital letters refer to matrices and bold lowercase letters represent vectors. The subscript on a lowercase letter $y_i$ represent the $i$-th element of vector $\mathbf y$. For two real number $a < b$, $[a,b]$ refers to the set of all real number $x$ satisfying $a<x<b$ while $[a,b]^n$ refers to the set of all $n$ dimensional vector with each element in $[a,b]$. $\mathbb{R}^n$ represents the space of all $n$ dimensional real vectors. $ \mathbb{C}^n$ represents the space of all $n$ dimensional complex vectors. Functions $real(\cdot): \mathbb{C}^n \rightarrow \mathbb{R}^{n}$ and $imag(\cdot): \mathbb{C}^n \rightarrow \mathbb{R}^{n}$  turn a complex vector to a real vector by taking the real or imaginary part of each element.

% Main contri + notation

\section{Deep Learning Basics}

A feedforward neural network \cite{goodfellow2016deep}, or multilayer perceptron, describes a mapping $f({\mathbf x}_i;{\mathbf w}_i): \mathbb{R}^{n_i} \rightarrow \mathbb{R}^{n_{i+1}} $in the $i$-th layer. Here the mapping is generally a linear transformation determined by the parameter ${\mathbf w}_i$ plus activation function to introduce nonlinearity. The output of the $i$-th layer is fed into next layer as input. The feedforward neural network with total layers of $L$ can be represented in (\ref{equ:mapping}).

\begin{equation}
\label{equ:mapping}
{\mathbf x}_{i+1} = f({\mathbf x}_i;{\mathbf w}_i), \quad i=1,2,\dots,L-1
\end{equation}

Given enough amount of training data, i.e. multiple pairs of input vector ${\mathbf x}_0$ and output vector ${\mathbf x}_L$, the mapping from input vector $x_0$ to output $x_L$ is approximated by the cascaded function in equation (\ref{equ:mapping}). Denote $f_i(\cdot) = f(\cdot;{\mathbf w}_i)$, we have ${\mathbf x}_L = f_{L-1}f_{L-2}\dots f_0({\mathbf x}_0)$. 

The training of a neural network is based on mini-batch stochastic gradient descent and back propagation methods. The gradient is calculated and propagated backward from the last layer to the first layer. The parameters ${\mathbf w}_i$ are updated accordingly.

\begin{figure}[!htbp]
  \centering
    \includegraphics[width=0.8\linewidth]{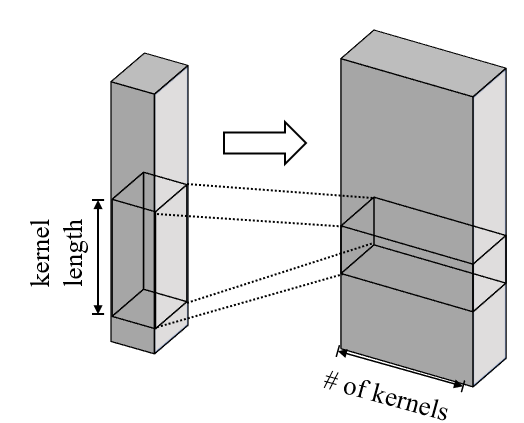}
  \caption{The illustraion of a 1D convolutiaonal neural network}
 \label{ill-cnn}
\end{figure}

There are multiple kinds of the layer $f(\cdot;{\mathbf w}_i)$. Common choices are fully connected layer, convolutional layer \cite{lecun1989generalization} and recurrent layer \cite{hochreiter1997long} etc. A fully connected layer is also called a dense layer, where each element of the input is connected to all the elements of its output vector. The dense layer is common and effective in most of the application areas, but is relatively high in computational complexity. The convolutional layer consists of a set of learnable filters called kernel. It performs a simple convolution along the input data. In our work, a 1D convolutional layer is frequently used to convolve the sequence with learnable filters. An illustration of how a convolutional layer works is shown in Fig. \ref{ill-cnn}. 

We also use a layer wrapper called time-distributed wrapper. It applies a layer to every temporal slice of its input instead of all the elements of the input. The function of a time-distributed dense on a 2D array is similar to a 1D convolutional layer with kernel size = 1.

\section{System Model}

\begin{figure}[!htbp]
  \centering
    \includegraphics[width=\linewidth]{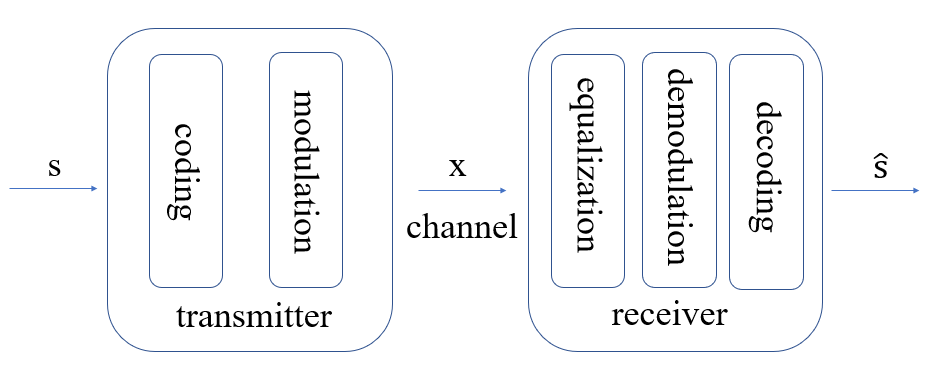}
  \caption{The block diagram for a typical communication system}
 \label{blockdiagram}
\end{figure}

We consider a typical communication system with block diagram in Fig. \ref{blockdiagram}. The input data  modeled as a sequence of i.i.d. bits ${\mathbf s}$ is coded and modulated by the transmitter. The transmitted data $x_i$ is transmitted through a linear time-invariant channel with impulse response coefficients $h_n$. The distortion and noise introduced by channel can be modeled as equation (\ref{equ:channel}).

\begin{equation}
\label{equ:channel}
y_i = \sum_{n=0}^{L_h-1}h_nx_{i-n} + n_i
\end{equation}

The received data ${\mathbf y}$ is equalized, demodulated and decoded by the receiver to recover the original sequence $\hat{{\mathbf s}}$. 

There is a restriction on the power of transmitted symbol in the real transmitter. Usually the average power for the signal is constrained to 1, i.e. $\frac{1}{n}||\mathbf{x}||_2 = 1 $.

The procedure of the communication system can be viewed as the cascade of three function $\hat{s} = g_3(g_2(g_1(s)))$. Here $g_1: [0, 1]^n \rightarrow \mathbb{C}^m$ is the function of the operation in transmitter. $g_2: \mathbb{C}^m \rightarrow \mathbb{C}^m $ is the channel function defined in equation (\ref{equ:channel}). $g_3:  \mathbb{C}^m \rightarrow [0, 1]^n $ is the receiver function to recover the original information.

The similarity between the representation of communication system and neural network leads us to optimize the communication system using neural networks. In the rest part of the paper, we study the problem of jointly optimizing function $g_1$ and $g_3$ with fixed $g_2$.

\section{ Autoencoder for Time Domain Transmission}

\subsection{Network Structure}

The property of neural network enables us to train the model using only input bit sequences under different channel state. 

\begin{figure}[!htbp]
  \centering
    \includegraphics[width=0.8\linewidth]{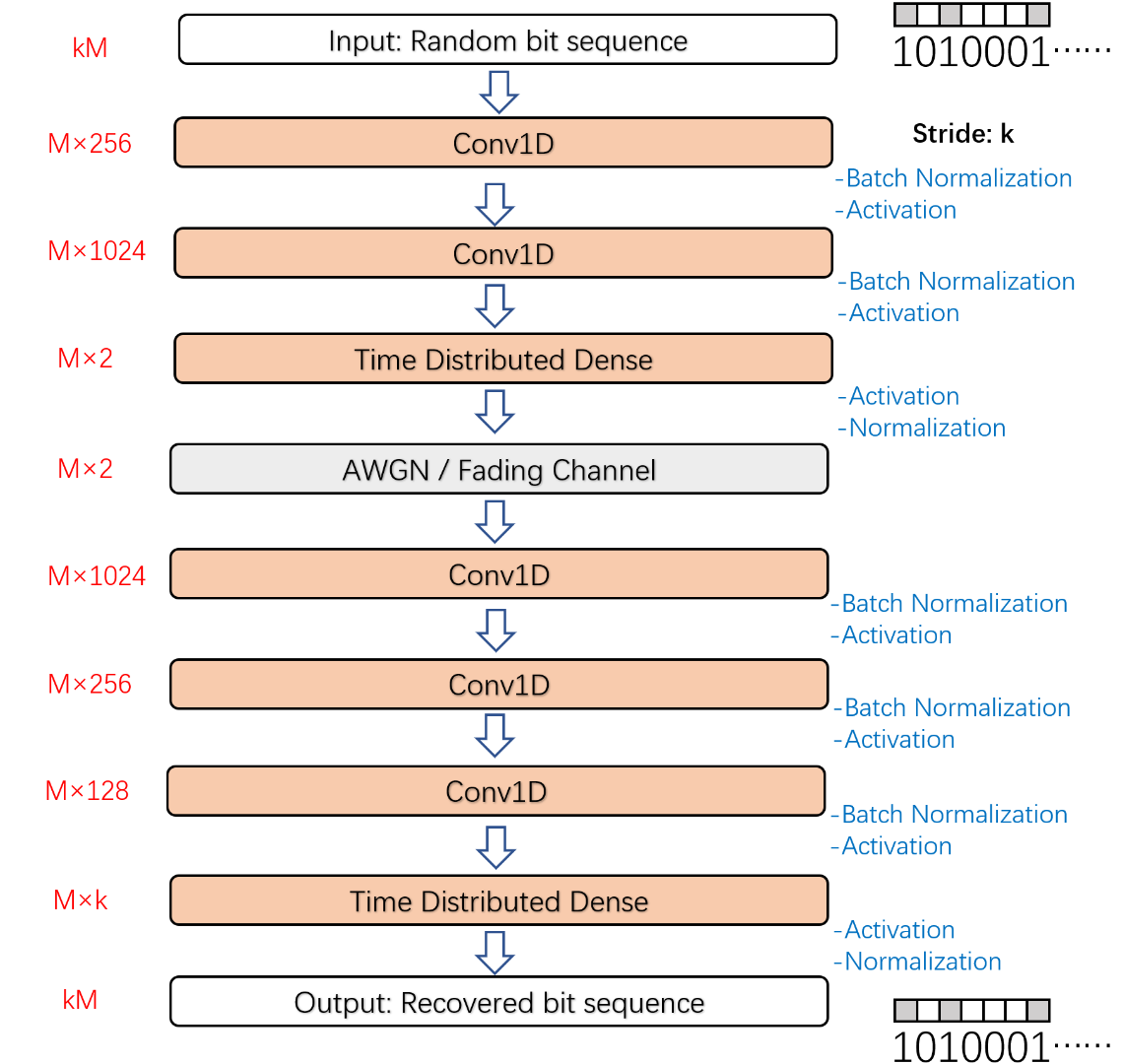}
  \caption{DNN framework for time domain transmission in physical layer. The input and output is marked with white box. The trainable layers are in orange and untrainable layers are in gray.}
 \label{dnnfortime}
\end{figure}

We jointly optimize the transmitter and receiver with an autoencoder structure. Convolutional neural network is used considering the sequential property of the input sequence. The network structure is shown in Fig. \ref{dnnfortime}.

Assume that the system is transferring $k \times M$ bits information $s \in \{0,1\}^{k \times M}$. Here $k$ represents the number of bits each symbol carries. And $M$ is the length of symbols. The transmitter maps the input bit stream to a sequential complex vector $x \in \mathbb{C}^M$ and transmits $x$ to the channel in the time domain. The received signal $y \in \mathbb{C}^N$ with distortion and noise is then equalized and demapped in the receiver to recover the original bit stream $s$. 

We compress the length of the input from $k \times M$ to $M$ in the first convolutional layer with stride size = $k$. And we combine the usage of time-distributed dense layer with convolutional layer to introduce further correlation and nonlinearity in the transmitter. The transmitted symbol $\mathbf{X} = [real(\mathbf{x});imag(\mathbf{x})]$ is of size $M \times 2$ as a complex vector. In the channel layer, the input symbols are first normalized to satisfy the power constraint. For AWGN channel, only additive white gaussian noise is added to the normalized symbols. For fading channels, the normalized symbols first convolve with the impulse response in the time domain. The convolution in complex number is implemented as in equation (\ref{equ:imagreal}). In neural networks, it can be represented by a 1D convolutional layer that convolves $\mathbf{X}$ with a 3D tensor.

\begin{equation}
\label{equ:imagreal}
\begin{aligned}
real(y_i) = &\sum_{n=0}^{L_h-1}(real(h_n)real(x_{i-n}) \\
            & - imag(h_n)imag(x_{i-n})) + real(n_i) \\
imag(y_i) = &\sum_{n=0}^{L_h-1}(real(h_n)imag(x_{i-n}) \\
            & - imag(h_n)real(x_{i-n})) + imag(n_i) \\
\end{aligned}
\end{equation}

Generally a neural network suffers from the restriction of input shape, i.e. the length of input for testing shall be the same as that in training procedure. However, due to the locally-connected property of convolutional layer and time-distributed layer, the proposed network structure is able to accept input sequences of any length without the need of retraining the whole model. Thus the system can process long sequences while trained on short ones.

We conduct massive experiments to analyze the performance of our model. We train our model separately on AWGN channel and fading channel. In the rest part of this chapter, we give thorough analysis on the result of the learned system.

\subsection{Setting}

In our experiment, we set $k=6$ and compare the learned system with 64QAM for AWGN channel, and 64QAM plus minimum mean square error (MMSE) estimation  \cite{johnson2004minimum} for fading channel. We also test the proposed model on $k=8$ case with 256QAM+MMSE in fading channel to prove the extensibility of our model. The modulation scheme is selected considering the throughput fairness. For the training and testing our model, we randomly generate i.i.d. bit sequences. The generated dataset is separated arbitrarily into training set, validation set and test set. The property of convolutional neural network enables the network to process input sequence of any length without changing network parameters. The change in the length of the input sequence would not affect the performance of our system. Thus we fix $M$ to be $400$.

We train the autoencoder with $30,000$ training samples and test that with $10,000$ test data. The learning rate is set to be $0.001$ and the batch size is $32$. We train the system at a specific signal to noise ratio (SNR) but test at a wide range of SNR, as well as robustness and adaptivity to deviations from the AWGN and fading setting. We would like to show that although the state space of input bit sequence is as large as $2^{k \times M}$, the model can generalize well with mere $30,000$ training samples.

\begin{figure}[!htbp]

\begin{minipage}[t]{0.51\linewidth}
\centering
\includegraphics[width=\linewidth]{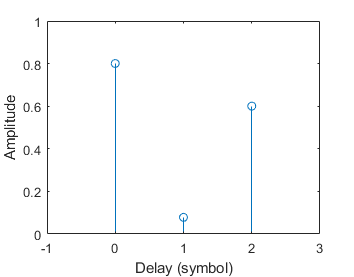}
%\caption{fig1}
%\label{fig:side:a}
\end{minipage}%
\begin{minipage}[t]{0.51\linewidth}
\centering
\includegraphics[width=\linewidth]{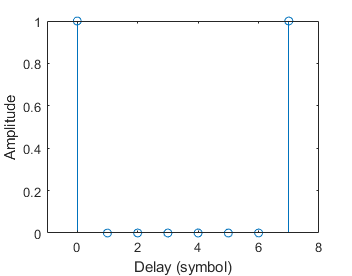}
%\caption{fig2}
%\label{fig:side:b}
\end{minipage}

\caption{The amplitude of two fading channels tested. The left is referred to as channel A and the right is channel B.}
 \label{fig:stem}
\end{figure}

In the time domain transmission system, we test three different channels, one AWGN channel and two fading channels. The amplitude and delay for two fading channels are plotted in Fig. \ref{fig:stem}. The integer in x-axis is the delay of each path in the unit of symbols. Note that the phase information is omitted in the figure. 

\subsection{AWGN Channel}

\begin{figure}[!htbp]
\centering
\includegraphics[width=0.8\linewidth]{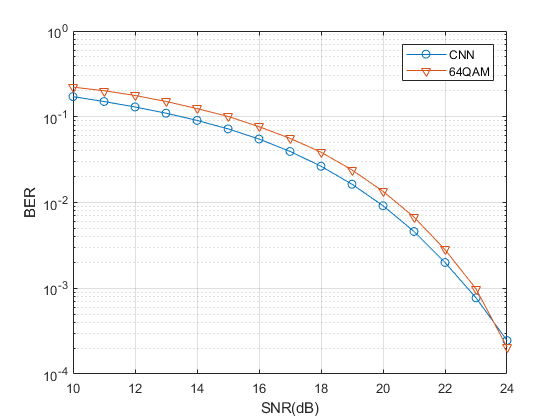}
\caption{The comparison between 64QAM and the trained system under AWGN channel. }
\label{fig:64awgn}
\end{figure}

The BER performance for AWGN channel is shown in Fig. \ref{fig:64awgn}. The CNN based autoencoder can bring up to 1 dB gain when bit error rate (BER) ranges from $10^{-1}$ to $10^{-3}$.

\begin{figure}[!htbp]
\centering
\includegraphics[width=0.8\linewidth]{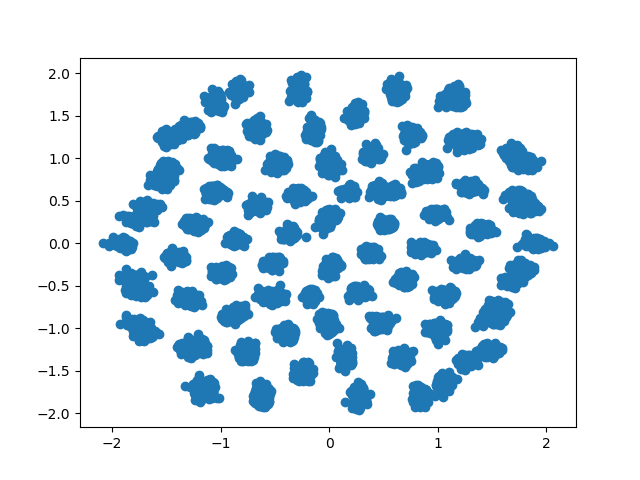}
\caption{The learned constellation diagram for AWGN channel. }
\label{fig:consdia}
\end{figure}

The constellation diagram is shown in Fig. \ref{fig:consdia}. We plot $40,000$ modulated symbols in a complex plane. The system learns an amplitude and phase-shift keying
 (APSK) like constellation method. However, the symbols seem not to concentrate into the finite constellation points, but are distributed as clusters instead. The clustered structure of constellation symbols may have higher requirements for antennas, but may bring extra gain for low SNR regime. The learned constellation is close to a Gaussian distribtion, which is optimal for a maximum likelihood receiver according to information theory.

 %% Add sth related to mutual information

\subsection{Fading Channel}

We train the system under two Rayleigh fading channels in Fig. \ref{fig:stem}. The benchmark system adopts 64QAM modulation and MMSE detection method. MMSE is assumed to have perfect channel state information.

\begin{figure}[!htbp]
\centering
\includegraphics[width=0.8\linewidth]{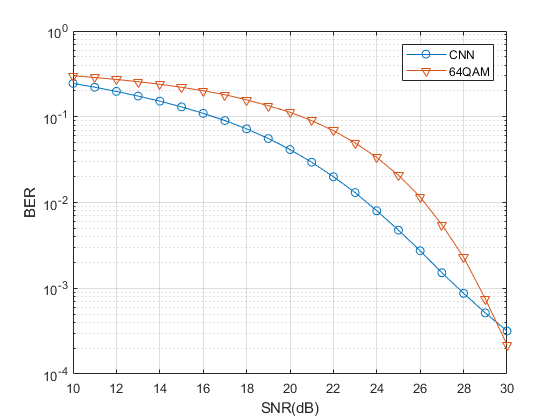}

\caption{ BER comparison between 64QAM+MMSE and the system under fading channel A. }

\label{fig:compchana}
\end{figure}

The comparison between CNN based method and 64QAM for fading channel A is shown in Fig. \ref{fig:compchana}.

\begin{figure}[!htbp]
  \centering
    \includegraphics[width=0.8\linewidth]{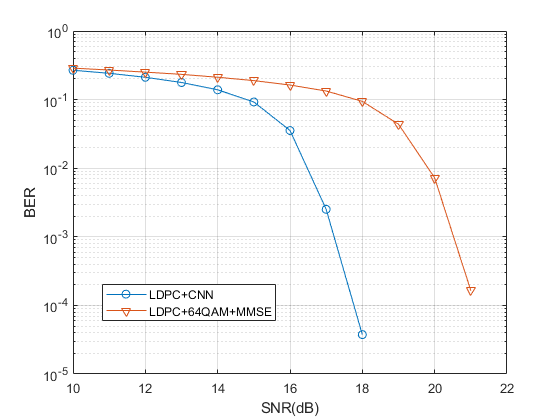}
  \caption{BER comparison after combining with LDPC under fading channel A.}
 \label{fig:ldpc}
\end{figure}

 The trained system is able to bring up to 4 dB gain. The system is outperformed by the benchmark in high SNR regime. This also happens in the case of AWGN channel. And this phenomenon is not related to the SNR used in training procedure. We would like to point out that the performance decay in high SNR would not affect the advantage of our method in practical system since our system can be easily combined with any coding module as long as the output sequence of the coding module is i.i.d., which is easy to satisfy with current interleaving technique. The output of the neural network is a real number in $[0,1]$. This can be viewed as a probability and fed into any soft-decoding system as soft input.  We add a standard LDPC coding system \cite{livshitz2012low} with code rate = $\frac{1}{2}$ to both 64QAM+MMSE system and our scheme. We can see in Fig. \ref{fig:ldpc} that after combining with a coding module, our system maintains a performance advantage over the entire SNR range.  
 
\begin{figure}[!htbp]
  \centering
     \includegraphics[width=0.8\linewidth]{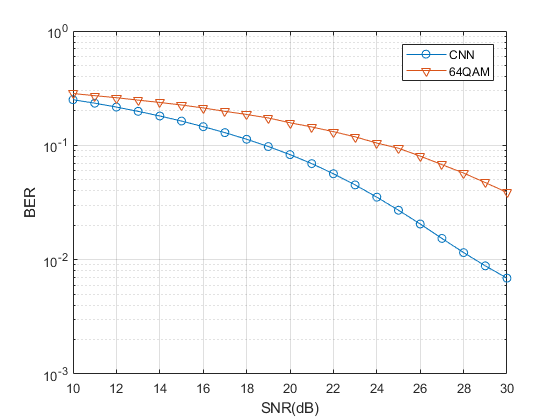}
  \caption{BER comparison between 64QAM+MMSE and the system under fading channel B.}
 \label{fig:2shutime}
\end{figure}

Furthermore, our system is able to bring higher gain for a channel with higher delay. We train the neural network under fading channel B. The result is shown in Fig. \ref{fig:2shutime}.

%The learned constellation diagram for fading channel is distributed all over the complex plane. 
Traditional 64QAM system is highly tailored for AWGN channel. For a known fading channel, the proposed neural network is able to introduce precoding in modulation procedure. The transmitter and receiver can be jointly adjusted to fit the channels, thus making the system more robust to fading.

\begin{figure}[!htbp]
  \centering
     \includegraphics[width=0.8\linewidth]{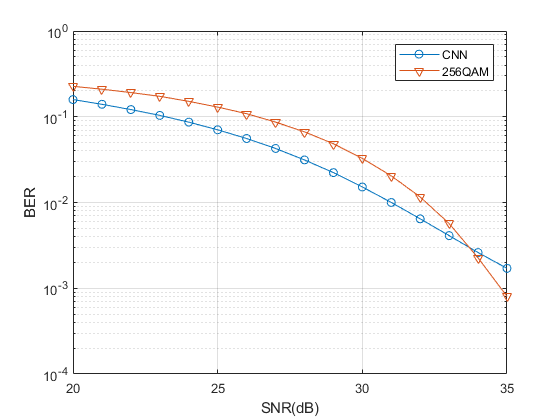}
  \caption{BER comparison between 256QAM+MMSE and the system under fading channel A.}
 \label{fig:256qam}
\end{figure}

The proposed structure can also be extended to different $k$. We train the neural network under $k=8$ and compare the result with 256QAM under channel A. The comparison is shown in Fig. \ref{fig:256qam}. Although the neural network is not as good as 256QAM in high SNR regime, the performance is guaranteed to maintain an advantage over the whole SNR regime after combining with coding system, as is the case shown above.

\subsection{Robustness}

\begin{figure}[!htbp]
  \centering
    \includegraphics[width=0.8\linewidth]{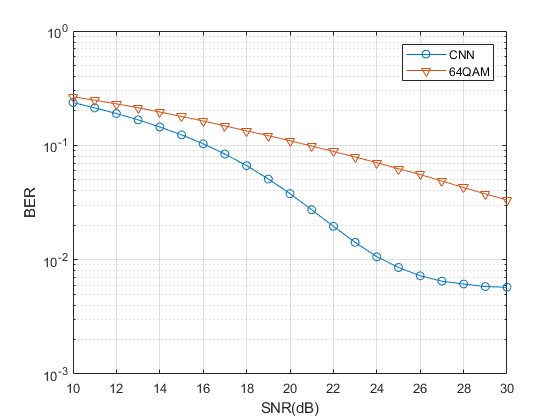}
  \caption{BER comparison between 64QAM and CNN for bursty AWGN channel.}
 \label{fig:bursty2}
\end{figure}

In this section, we show the neural network system is much more robust against the variations in the channel. This makes, among other things, the proposed system much more attractive alternative to traditional modulation methods in practice, where the channel model is not available. We consider two kinds of channel variations. One of them is the introduction of non-Gaussian noise. We test bursty AWGN channel where a small fraction of noise has much higher variance than other on the neural network system trained under AWGN channel, and compare that with 64QAM. Usually bursty AWGN channel is used to model inter-cell interference in OFDM cellular systems or co-channel radar interference \cite{Fertonani2008OnRC}. The comparison result is shown in Fig. \ref{fig:bursty2}. The result is expected since convolutional neural network would introduce correlation in a longer range of symbols than 64QAM, making system more robust to burst noise.

\begin{figure}[!htbp]
\subfigure[]{
\begin{minipage}[t]{\linewidth}
\centering
\includegraphics[width=0.8\linewidth]{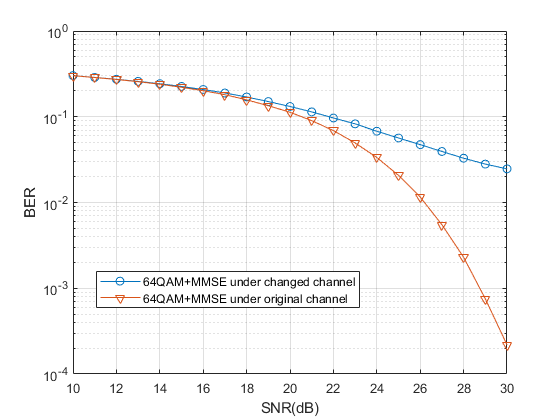}
%\caption{fig1}
%\label{fig:side:a}
\end{minipage}%
}

\subfigure[]{
\begin{minipage}[t]{\linewidth}
\centering
\includegraphics[width=0.8\linewidth]{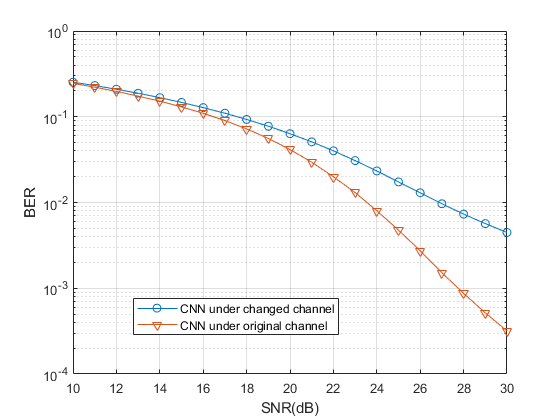}
%\caption{fig2}
%\label{fig:side:b}
\end{minipage}
}
\caption{The comparison of robustness of 64QAM+MMSE (a) and neural network (b). Both methods are tested under original channel and changed channel.}
 \label{fig:robustness}
\end{figure}

Another kind of channel imperfections is related to the time-variant property of real channel and inaccurate channel estimation. Channel in real world may change within a single frame. The signal may go through a slightly different channel from the one previous pilot passes. Furthermore, channel estimation cannot ensure perfect recovery of channel state information. In our experiment, we train the model on a fixed channel A. And test it with a different channel that adds a white Gaussian white noise with standard deviation = 0.05 to channel A. For MMSE detection, it takes the fixed channel A as channel state information. We test both 64QAM+MMSE and neural network system 100 times and take the average BER performance. The comparison of robustness between 64QAM+MMSE and neural network is shown in Fig. \ref{fig:robustness}. The channel variation affects 64QAM+MMSE more significantly than neural network structure.

 From the experiments above, we can see that the proposed method is especially promising under extreme channel condition, like severe fading, or frequent bursty noise. Furthermore, it is more robust to channel variation than traditional methods.
 
\subsection{Time Complexity Comparison}

We provide both simulation and numerical analysis for analyzing time complexity. We first test the time complexity by running demodulation plus detection algorithms and the receiver part in neural network on a Intel (R) Corel (TM)
i7-7700HQ CPU @ 2.80GHz CPU and an NVIDIA GeForce GTX 1060 GPU. The platform in this experiment is python+keras \cite{chollet2015keras}. For AWGN channel, only demodulation is needed to recover the bit sequence. For fading channel, MMSE is also included in the receiver, which takes long for the FFT step. We test both methods for 100 sets of data and take the average. The comparison is shown in Table.\ref{tab:time}. 

\begin{table}[!htbp]
\caption{Time Complexity Comparison between traditional demodulation method and neural network.}
\centering
\label{tab:time}
%\wuhao[1.5]
\begin{tabular}{c|c|c}
\toprule[2pt]
\diagbox{Channel}{Method} & 64QAM(+MMSE) & CNN  \\
\hline

%\midrule[1pt]
AWGN & \bf 1.501s & 3.695s \\
Fading & 8.709s &\bf 3.751s\\

\bottomrule[1pt]
\end{tabular}
\end{table}

For transmitted bit sequence with length $n$, the time complexity for both CNN and QAM demodulation is $O(n)$. However, for MMSE detection, FFT requires a $O(nlogn)$. CNN is able to substitute the demodulation plus detection algorithm with a lower time complexity in theory and higher accuracy. Thus CNN based framework is quite suitable for designing communication system in fading channels.

With the development of neural network, the dimension of the CNN can be potentially reduced with techniques such as network pruning and distillation. Parallelization is also possible in the multiplicative units in the neural network, as well as pipelining. Neural network can be further accelerated by specially designed hardware framework like GPU, FPGA and TPU etc. These designs along with a careful analysis of the fixed point arithmetic requirements of the different weights are under active research. The efficiency of neural network can be further improved in the future.

\section{Autoencoder for Frequency Domain Transmission}

\subsection{Network Structure}
In the previous chapter, we studied the effect of the multipath fading channel. In Orthogonal Frequency Division Multiplexing (OFDM) \cite{le1995coded} system, the inter-symbol interference can be eliminated by introducing guard interval between each subcarrier. Cyclic prefix is a typical guard interval, which makes each subcarrier orthogonal to each other. In the following part of this chapter, we assume perfect cyclic prefix as guard interval. Thus there is no inter-symbol inference and inter-channel inference. However, there is still fading on each subcarrier. One of the most common equalization method for this issue is zero forcing (ZF) \cite{peel2005vector}. Since the fading on individual subcarrier might be quite significant, it is hard to recover the signal from those subcarriers due to poor SNR. The information carried by some subcarriers might experience deep fading and thus get lost during transmission. In fact, a zero order equalization system would cause inevitable loss of information \cite{daubechies1986painless}.  By introducing correlation between subcarriers, the information can be carried by nearby subcarriers. Thus the burst error on subcarriers with deep fading can be decreased.

\begin{figure}[!htbp]
  \centering
    \includegraphics[width=0.8\linewidth]{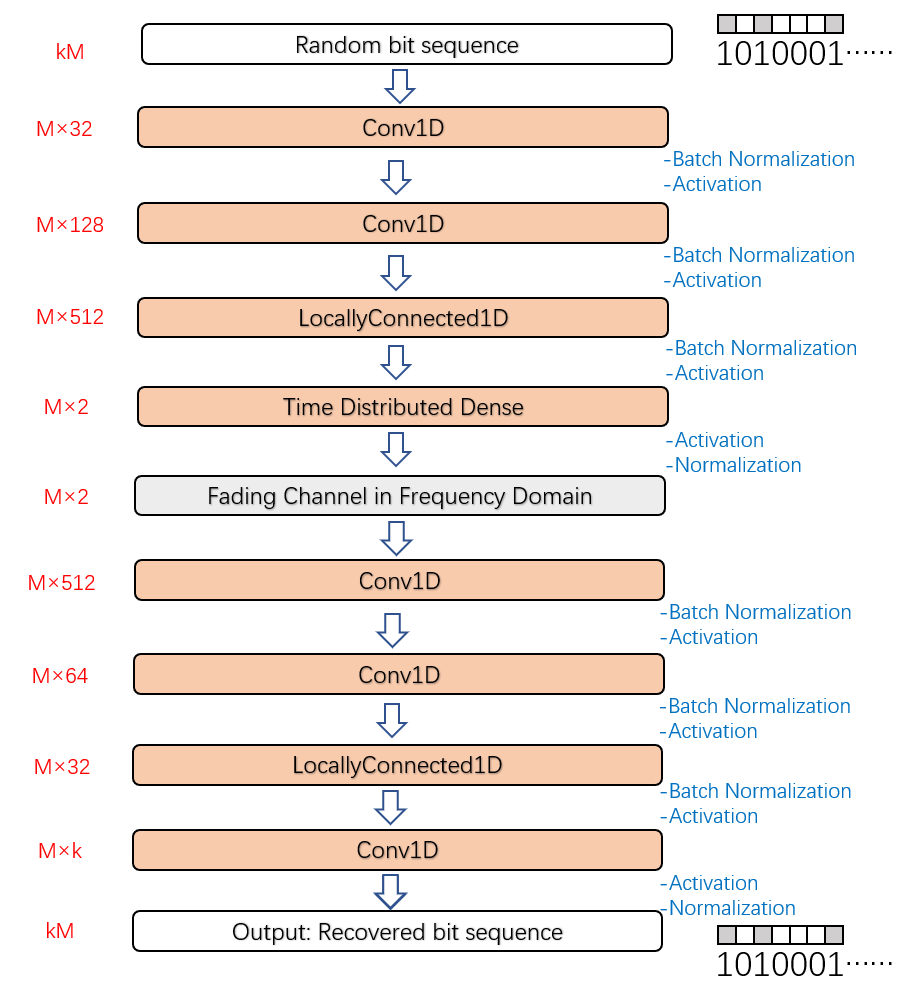}
  \caption{DNN framework for frequency domain transmission in physical layer.}
 \label{fig:dnnforfreq}
\end{figure}

Based on previous network structure, we design a frequency domain equalization system that is able to retrieve more information than ZF. Since the fading on each subcarrier is different, a convolutional layer that shares weight along the whole input sequence may not be suitable. Here locally connected layer is used to substitute some of the convolutional layers in previous structure. The locally connected layer works similarly as the convolutional layer, except that weights of kernels are unshared. That is, a different set of filters is applied at each different patch of the input. The whole network structure is shown in Fig. \ref{fig:dnnforfreq}.

\subsection{Simulation}
\begin{figure}[!htbp]
  \centering
    \includegraphics[width=0.8\linewidth]{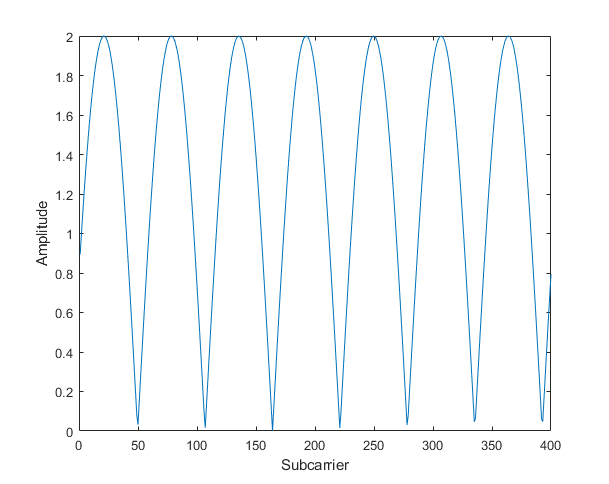}
  \caption{The amplitude of frequency selective fading for each subcarrier in frequency domain.}
 \label{fig:ofdmchan}
\end{figure}

We test the new structure under the case of an OFDM system transmission. The bit sequence is modulated and transmitted in frequency domain. All other settings are the same as the previous example. We consider the frequency transformation of channel B in Fig. \ref{fig:stem}. The frequency selective fading channel in frequency domain is shown in Fig. \ref{fig:ofdmchan}. We train and test our model on the same channel. And compare that with 64QAM+ZF method. Here ZF is assumed to have accurate channel state information. But
the zero points in the fading plot would prohibit a significant number of subcarriers from transmitting information to the receiver.

\begin{figure}[!htbp]
  \centering
    \includegraphics[width=0.8\linewidth]{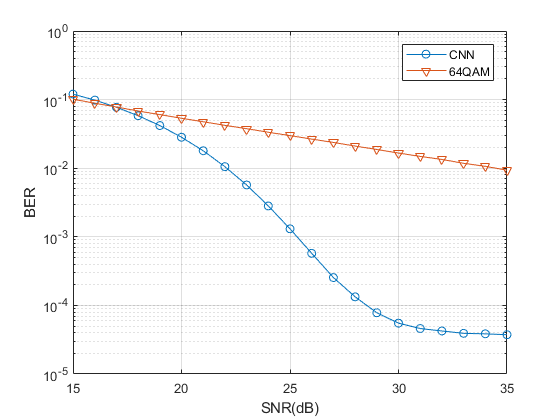}
  \caption{BER comparison between 64QAM+ZF and CNN under OFDM frequency selective fading channel.}
 \label{fig:ofdmcomp}
\end{figure}

The BER performance for both 64QAM+ZF and CNN under OFDM frequency selective fading channel is shown in Fig. \ref{fig:ofdmcomp}. The BER for CNN based system can reach as low as $10^{-4}$ while the BER for traditional 64QAM+ZF method goes down slow. Because of the correlation introduced in convolutional layer, the neural network is able to assign different amount of information to different subcarriers according to the statistical property of the noise. The function of the neural network is similar to adaptive bit loading techniques \cite{barreto2001adaptive} that assign bits efficiently based on subcarrier quality. This may explain the huge gain the neural network brings compared to traditional 64QAM+ZF system.

\section{Conclusion}

In this paper, we propose a convolutional autoencoder structure that is able to automatically design communication physical layer scheme according to different channel status. The system has no restriction on the length of input bit sequence. We conduct massive experiment to give empirical evidence for the superiority of the proposed system. The neural network has lower time complexity and higher accuracy especially for fading channel, and is also quite robust to channel variation. The framework can also be extended to OFDM system which transmits in frequency domain.

The trained autoencoder may not compete with the state-of-the-art system that is optimized over the past decades. But the autoencoder is able to learn the way of mapping and demapping for any known channel without prior mathematic model and analysis. We may further explore the feasibility and utility of neural network based communication methods in following aspects.

\begin{itemize}
\item One of the most important goals of designing a communication system is to maximize the capacity, i.e. the mutual information between input and output. However, since the constellation diagram in neural network based system is continuously distributed in the complex plane. It is hard for us to estimate the mutual information accurately, let alone optimizing the mutual information within neural network. A framework for analyzing mutual information in neural network based communication system may significantly enlarge our knowledge about both neural network and communication system.
\item An iterative, soft-input soft-output receiver can significantly improve the BER performance. Designing a receiver with both log likelihood ratio and received symbol as input using neural network may also enable iteration inside receiver, thus improving the current performance. 
\item The performance of neural network is still not as good in high SNR regime. It is important to figure out how autoencoder can learn a communication physical layer that outperforms existing communication techniques even in the high-SNR regions.

\end{itemize}

\bibliographystyle{IEEEtran}
\bibliography{ref}

\end{document}